\documentclass[prd,superscriptaddress,showpacs,twocolumn]{revtex4}

\usepackage{graphicx}%
\usepackage{dcolumn}
\usepackage{amsmath}

\usepackage[dvips]{epsfig}
\usepackage{amssymb}
\usepackage{array,longtable}
\usepackage{amscd}

\begin{document}



\newcommand{\be}{\begin{eqnarray}}
\newcommand{\ee}{\end{eqnarray}}
\newcommand{\bse}{\begin{subequations}}
\newcommand{\ese}{\end{subequations}}

\newcommand{\bs}{\boldsymbol}
\newcommand{\mbb}{\mathbb}
\newcommand{\mcal}{\mathcal}
\newcommand{\mfr}{\mathfrak}
\newcommand{\mrm}{\mathrm}

\newcommand{\ovl}{\overline}
\newcommand{\p}{\partial}
\newcommand{\f}{\frac}
\newcommand{\diff}{\mrm{d}}
\newcommand{\lan}{\left\langle}
\newcommand{\ran}{\right\rangle}

\newcommand{\ga}{\alpha}
\newcommand{\gb}{\beta}
\newcommand{\gc}{\gamma}
\newcommand{\Gd}{\Delta}
\newcommand{\gd}{\delta}
\newcommand{\Gc}{\Gamma}
\newcommand{\gl}{\lambda}
\newcommand{\Gl}{\Lambda}
\newcommand{\gk}{\kappa}
\newcommand{\go}{\omega}
\newcommand{\Go}{\Omega}
\newcommand{\Gs}{\Sigma}
\newcommand{\gs}{\sigma}
\newcommand{\veps}{\varepsilon}
\newcommand{\eps}{\epsilon}
\newcommand{\Gt}{\Theta}

\newcommand{\sn}{\mrm{sn}}
\newcommand{\cn}{\mrm{cn}}
\newcommand{\dn}{\mrm{dn}}
\newcommand{\am}{\mrm{am}}
\newcommand{\sech}{\mrm{sech}}
\newcommand{\sign}{\mrm{sign}}
\newcommand{\artanh}{\mrm{artanh}\,}

\newcommand{\csp}{\;,\qquad\qquad}
\newcommand{\fa}{\forall\;}

\newcommand{\N}{\mbb{N}}
\newcommand{\R}{\mbb{R}}
\newcommand{\D}{\mcal{D}}
\newcommand{\C}{\mcal{C}}
\newcommand{\Nn}{\mcal{N}}
\newcommand{\V}{\mcal{V}}
\newcommand{\T}{\mcal{T}}

\newcommand{\im}{\mrm{image}\;}
\newcommand{\num}{\mrm{\#}}

\newcommand{\kB}{k_\mrm{B}}


\title{Relativistic diffusion processes and random walk models} 
\author{J\"orn Dunkel}
\email{joern.dunkel@physik.uni-augsburg.de}
\author{Peter Talkner}
\author{Peter H\"anggi}
\affiliation{Institut f\"ur Physik, Universit\"at Augsburg,
 Theoretische Physik I,  Universit\"atstra{\ss}e 1, D-86135 Augsburg, Germany}

\date{\today}

\begin{abstract}
The nonrelativistic standard model for a continuous, one-parameter diffusion process in position space is the Wiener process. As well-known, the Gaussian transition probability density function (PDF) of this process is in conflict with special relativity, as it permits particles to propagate faster than the speed of light. A frequently considered alternative is provided by the telegraph equation, whose solutions avoid superluminal propagation speeds but suffer from singular (non-continuous) diffusion fronts on the light cone, which are unlikely to exist for massive particles. It is therefore advisable to explore other alternatives as well. In this paper, a generalized Wiener process is proposed that is continuous, avoids superluminal propagation, and reduces to the standard Wiener process in the non-relativistic limit. The corresponding relativistic diffusion propagator is obtained directly from the nonrelativistic Wiener propagator, by rewriting the latter in terms of an integral over actions. The resulting relativistic process is non-Markovian, in accordance with the known fact that nontrivial continuous, relativistic Markov processes in position space cannot exist. Hence, the proposed process defines a consistent relativistic diffusion model for massive particles and provides a viable alternative to the solutions of the telegraph equation. 
\end{abstract}


\maketitle

\section{Introduction}
\label{introduction}

It is well-known for a long time that the nonrelativistic diffusion equation (DE)~\cite{Be67,Gardiner,HaTo82,Roepstorff}
\be\label{e:diffusion}
\f{\p}{\p t} \rho(t,x)=D\,\f{\p^2}{\p x^2} \rho(t,x),\qquad 
t\ge 0,\; D>0
\ee
is in conflict with the postulates of special relativity. A simple way to see this is to consider the propagator of Eq.~\eqref{e:diffusion}, which for $d=1$ space dimensions, is given by
\be\label{e:solution_non-rel}
p(t,x|x_0)=
\f{\Theta(t)}{\left(4\pi Dt\right)^{1/2}}\,
\exp\biggl[-\f{(x-x_0)^2}{4D t}\biggr]
\ee
[Heaviside's theta-function is defined by \mbox{$\Gt(t)=1,t\ge 0$} and \mbox{$\Gt(t)=0,t< 0$}]. This propagator~\eqref{e:solution_non-rel} represents the solution of Eq.~\eqref{e:diffusion} for the initial condition \mbox{$\rho(0,x)=\gd(x-x_0)$}. That is, if $X(t)$ denotes the path of a particle with fixed initial position $X(0)=x_0$, then $p(t,x|x_0)\diff x$ gives the probability that the particle is found in the infinitesimal volume element $[x,x+\diff x]$ at time~$t>0$. As evident from Eq.~\eqref{e:solution_non-rel}, for each $t>0$ there is a small, but non-vanishing probability that the particle may be found at distances $|x-x_0|>ct$, where $c$ is the speed of light.
\par
This problem has attracted considerable interest during the past 100 years~(see, e.g., \cite{Go50,Ru57,Sc61,Ha65,Du65,MaWe96,1997DeMaRi,1998DeRi,BoPoMa99,HePa01,KoLi00,KoLi01,AbGa04,AbGa05,Ra05,OrHo05,DuHa05a,DuHa07,DuHa06b,2004De,Ko05,FrLJ06} and references therein). Nonetheless, it seems fair to say that a commonly accepted solution is still outstanding. Apart from the profound theoretical challenge of developing a consistent relativistic diffusion theory, several practical applications exist, e.g., the analysis of data from high energy collision experiments~\cite{Wo04,AbGa04,AbGa05,2005RaGrHe,2006PhRvC..73c4913V}. In this context, an often considered alternative to Eq.~\eqref{e:diffusion} is given by the telegraph equation~(TE)~\cite{Ta22,Go50,RevModPhys.61.41,RevModPhys.62.375,MaWe96,KoLi00,KoLi01,Ko05,AbGa04,AbGa05}
\be\label{e:telegraph}
\tau_d \f{\p^2}{\p t^2}\rho + \f{\p}{\p t} \rho=D\,\nabla^2 \rho,
\ee 
where $\tau_d$ is an additional relaxation time parameter.  As e.g. discussed by Masoliver and Weiss \cite{MaWe96} or Abdel-Aziz and Gavin~\cite{AbGa04}, the diffusion fronts described by Eq.~\eqref{e:telegraph} spread at finite absolute speed $c_d=\pm (D/\tau_d)^{1/2}$; cf. Fig.~1 of Ref.~\cite{AbGa04}. In particular, by fixing $c_d=c$, the solution of Eq.~\eqref{e:telegraph} defines a simple   relativistic, \emph{non-Markovian}~\footnote{Because of the second order time derivative, the TE~\eqref{e:telegraph} describes a \emph{non-Markovian} diffusion process. A general non-existence theorem for Lorentz-invariant Markov processes in position space can be found in Dudley's seminal paper~\cite{Du65}.} diffusion process without superluminal propagation. The \lq nonrelativistic limit\rq\space corresponds to letting $c_d\to \infty$ in Eq.~\eqref{e:telegraph}, which gives $\tau_d\to 0$ and thus leads back to Eq.~\eqref{e:diffusion}.
\par 
However, apart from these appealing properties the solutions of Eq.~\eqref{e:telegraph} exhibit a peculiar feature that is rather unexpected from a simple diffusion theory: The diffusion fronts are given by running $\gd$-peaks whose intensity decreases as time grows; cf.~Eq.~(32) in Ref.~\cite{MaWe96}. The appearance of such singularities can be understood by the fact that the one-dimensional TE~\eqref{e:telegraph} can be derived from a so-called \emph{persistent} random walk (RW) model~\cite{Fu17,Fu22,Ta22,Go50,BoPoMa98}. 
This model assumes that a particle moves with constant absolute velocity between neighboring lattice points. At each lattice point, the particle is either back-scattered or transmitted, with the transmission probability being larger than the back-scattering probability (\emph{persistence}). As Goldstein~\cite{Go50} showed for the one-dimensional case, the continuum limit of this model leads to the TE (see Bogu\~n\'a et al.~\cite{BoPoMa98} for a generalization to higher space dimensions). 
\par
Hence, in contrast to the ordinary DE~\eqref{e:diffusion}, the TE~\eqref{e:telegraph} relies on asymmetric transition probabilities, which among others lead to a non-vanishing probability concentration at the diffusion fronts. From the practical point of view, the concept of persistent diffusion can certainly be useful in some situations, e.g., if one aims at describing the propagation of photons in thin slabs or foams~\cite{BoPoMa99,miri:031102, miri:031111,miri:031115}. In general, however, it seems unlikely that the relativistic generalization of the simple nonrelativistic diffusion theory would require conceptual modifications as severe as the introduction of persistence. In particular, for massive particles $\gd$-peaked diffusion fronts are unlikely to occur in nature, because such fronts would imply that a finite fraction of particles carries a huge amount of kinetic energy (much larger than $mc^2$). One may, therefore, wonder if it is possible to construct alternative, non-persistent diffusion processes, which are consistent with the basic requirements of special relativity, and converge to the solutions of the nonrelativistic DE~\eqref{e:diffusion} in a suitable limit case. This is the question the present paper aims to deal with.
\par
Before outlining our approach, a general remark might be in order. Usually, a diffusion theory intends to provide a simplified phenomenological description for the complex stochastic motion of a particle in a background medium (e.g., on a substrate~\cite{Fu17,Fu22,Ta22,Go50,BoPoMa98} or in a heat bath~\cite{DuHa07}). Thus, there exists a \emph{preferred} frame, corresponding to the rest frame of the substrate (or, more generally, the center-of-mass frame of the interaction sources causing the stochastic motion). It is therefore not expedient to look for Lorentz or Poincar\'e invariant spatial diffusion processes (cf. Sec.~5 of Montesinos and Rovelli~\cite{MoRo01}). Accordingly, we focus here on discussing simple diffusion models that comply with the basic requirements of special relativity in the rest frame of the substrate.
\par
Very often, diffusion models are based on time evolution equations like the DE~\eqref{e:diffusion}, derived either from an underlying microscopic model or as an approximation to a more complicated phenomenological model~\cite{Roepstorff,Gardiner,LaRoTi82,HaTaBo90,Kramers,Be67}. Unfortunately, apart from the TE~\eqref{e:telegraph}, the construction of relativistic DEs turns out to be very cumbersome. Among other things, this can be attributed to the fact that standard limiting procedures, as those leading  from discrete RW models~\cite{Gardiner} to Eq.~\eqref{e:diffusion}, typically result in superluminal propagation (cf. the discussion in Sec.~\ref{massive_model} below). 
Therefore,  a different approach is pursued in the present paper: Instead of aiming at a relativistic evolution equation in the first place, we try to obtain a continuous, relativistic generalization of Eq.~\eqref{e:solution_non-rel} by focussing on the structure of diffusion propagators.
\par
The paper is organized as follows. First, some basic properties of Markovian diffusion models for massive particles are reviewed (Sec.~\ref{massive_model}). In particular, it will be demonstrated in detail why it is impossible to construct a continuous relativistic Markov model in position space (Sec.~\ref{relativistic}). Subsequently, we propose a simple non-Markovian generalization of the nonrelativistic Wiener process (Sec.~\ref{axiomatic}). The specific functional shape of the relativistic propagator arises naturally, if one rewrites the Wiener propagator~\eqref{e:solution_non-rel} in terms of an action integral. The paper concludes with a summary of the main results.

\section{Markovian diffusion models}
\label{massive_model}

In this section, we consider a standard Markovian RW model that provides the basis for the subsequent considerations. First, the underlying equations and assumptions are summarized. Then it is discussed why the discrete RW model may lead to a useful nonrelativistic continuum model, but fails to do so in the relativistic case. For clarity, we focus on the one-dimensional case $d=1$ in this part~\footnote{The one-dimensional model described in this section may be easily generalized to higher space dimensions by replacing position/velocity coordinates $(X,V)$ with vectors $(\bs X,\bs V)$.}. 

\subsection{Basic equations and assumptions}
Consider a massive particle located at some known position $X(0)=x_0$ at the initial time $t_0=0$. Assuming that the particle performs random motions, we are interested in the distribution of the particle position $X(t)=x$ at time $t>0$. To obtain a simple RW model for the stochastic dynamics of the particle, we start from the formal identity 
\be\label{e:basic_equation}
X(t)=x_0+\int_0^t\diff s\; V(s),
\ee
where $V(s)$ is the velocity. Next we assume that the motion of the particle can be viewed as a sequence of free ballistic flights, interrupted by $(N-1)$ scattering events. Mathematically, this idea is formalized by introducing the discretization $\T_N=\{t_0,\ldots,t_N\}$ of the interval $[0,t]$, such that
$$
0=t_0<t_1<\ldots t_{N-1}<t_N=t.
$$ 
Defining the times-of-free-flight by
$$
\tau_i:=t_i-t_{i-1},\qquad i=1,\ldots N,
$$
we may discretize Eq.~\eqref{e:basic_equation} by using the approximation
\bse\label{e:basic_equation_2}
\be\label{e:basic_equation_2a}
X(t)= x_0+\sum_{i=1}^{N} \tau_i\,v_i.
\ee
Here,  $v_i=V(t_i)$ is the constant velocity of the particle between $t_{i-1}$ and $t_i$, i.e., before the $i$th collision. 
\par
Within the standard Markovian approach, it is commonly assumed that the velocities $(v_1,v_2,\ldots)$ can be viewed as independently distributed random variables with a given PDF 
\be
f_N(v_1,\ldots,v_N;\{\tau_i\})=\prod_{i=1}^N f(v_i;\tau_i).
\ee
For any fixed partition $\{\tau_i\}_{i=1,\ldots,N}$, this uniquely  determines the transition PDF for the position coordinate as 
\be\label{e:propagator}
p_N(t,x|x_0)=
\left[\prod_{i=1}^N\int \diff v_i\, f(v_i;\tau_i)\right]
\gd(x-X(t)),
\quad
\ee 
\ese
where $X(t)$ is the discretized path from Eq.~\eqref{e:basic_equation_2a}. Throughout, we abbreviate $p_N(t,x|x_0)=p_N(t,x|t_0,x_0)$ if the initial time is fixed as $t_0=0$.
\par
The continuum limit is recovered by letting $N\to\infty$, such that $\tau_i\to 0$ and
\bse\label{e:basic_equation_3}
\be
\sum_{i=1}^N\tau_i=t-t_0=t.
\ee
The transition PDF in the continuum limit can then be expressed as
\be\label{e:basic_equation_3b}
p(t,x|x_0)
=\lim_{N\to\infty} p_N(t,x|x_0).
\ee
\ese
Yet, the continuous Markov model~\eqref{e:basic_equation_3} is useful only if this continuum limit yields a nontrivial result. As a familiar example, we will recall how the (marginal) velocity distribution $f(v;\tau_i)$ is to be chosen in order to recover the nonrelativistic diffusion propagator~\eqref{e:solution_non-rel} from Eqs.~\eqref{e:basic_equation_2} and~\eqref{e:basic_equation_3}. Thereby, it will become evident why this Markovian approach fails to produce a useful continuum model in the relativistic case.

\subsection{Nonrelativistic diffusion model}
\label{nonrelativistic}
In the nonrelativistic case, we choose
\be\label{e:Maxwell}
f(v_i;\tau_i) 
= 
\left(\f{\tau_i}{4\pi D}\right)^{1/2} 
\exp\biggl(-\f{\tau_i\; v_i^2}{4D}\biggr).
\ee 
This can be interpreted as a Maxwellian velocity distribution with an inverse (effective) temperature 
\be\label{e:temperature}
\gb_i=(\kB T_i)^{-1}=\f{\tau_i}{2mD}
\ee
where $m$ denotes the particle rest mass, and $\kB$ the Boltzmann constant. Inserting Eq.~\eqref{e:Maxwell} into Eq.~\eqref{e:basic_equation_3b} gives
\be
p(t,x|x_0)
&=&\notag
\lim_{N\to\infty}\left[
\prod_{i=1}^N\int \diff v_i\,\left(\f{\tau_i}{4\pi D}\right)^{1/2}  \right]
\times\\
&& \notag\qquad
\exp\biggl(-\sum_{i=1}^N\f{\tau_i\, v_i^2}{4D}\biggr)
\times\\
&&\qquad\label{e:propagator_discrete}
\gd\biggl((x-x_0)-\sum_{i=1}^{N} \tau_i\,v_i\biggr).
\ee
By use of the identity
\be\notag
\gd(x)=\f{1}{2\pi}\int_{-\infty}^\infty \diff k\;e^{ikx}
\ee
it is straightforward to show that the rhs. of Eq.~\eqref{e:propagator_discrete} is equal to the nonrelativistic diffusion propagator from Eq.~\eqref{e:solution_non-rel}, and, hence, defines a Wiener process~\cite{Wi23}. Then, fixing the element of the functional integration by
\be
\int\D[V]=\lim_{N\to\infty}\left[
\prod_{i=1}^N\int \diff v_i\left(\f{\tau_i}{4\pi D}\right)^{1/2} \right]
\ee
one finds the following functional integral representation of the nonrelativistic diffusion propagator
\be\label{e:nonrelativistic_propagator_pi}
p(t,x|x_0)&=&\notag
\int\D[V]\;\exp\biggl[-\int_0^t\diff s\; \f{V(s)^2}{4D}\biggr]
\times\\
&&\qquad
\gd\biggl((x-x_0)-\int_0^t\diff s\; V(s)\biggr).
\qquad
\ee
Very often, path integral representations of diffusion propagators~\cite{LaRoTi82,Kleinert} are formulated in terms of curves $X(t)$ in position space, whereas Eq.~\eqref{e:nonrelativistic_propagator_pi} expresses the diffusion propagator~\eqref{e:solution_non-rel} by means of a functional integral over the space of the velocity curves. This is more convenient with regard to the subsequent discussion.
\par
\begin{figure}[t]
\epsfig{file=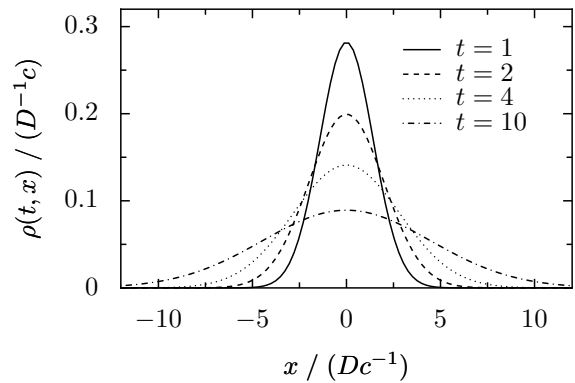}
\caption{Spreading of the Gaussian PDF $\rho(t,x)=p(t,x|0)$ from Eq.~\eqref{e:solution_non-rel} at different times $t$, where $t$ is measured in units of $D/c^2$. At initial time $t=0$, the PDF corresponds to a $\gd$-function centered at the origin.
\label{fig01} }
\end{figure}

The nonrelativistic PDF~(\ref{e:solution_non-rel},\ref{e:nonrelativistic_propagator_pi}) is depicted in~Fig.~\ref{fig01}. For later use, we mention that the mean square displacement of the Wiener process is given by~\cite{Be67} 
$$
\lan [X(t)-x_0]^2\ran:=
\int \diff x\;(x-x_0)^2\;p(t,x|x_0)=
2D\,t.
$$
With regard to the subsequent discussion, however, it is important to notice that the continuum limit $\tau_i\to 0$ implies that the effective temperature $T_i$ of the Maxwell distribution~\eqref{e:Maxwell} diverges, $T_i\to\infty$; i.e.,  the second moment $\lan v_i^2\ran$ of the velocity distribution $f(v_i;\tau_i)$ tends to infinity in the continuum limit. As discussed in the next part, this divergence is essential for ensuring that those scenarios, where a particle leaves its initial position, retain a non-vanishing probability in the continuum limit.

\subsection{Relativistic case}
\label{relativistic}

The superluminal propagation of particles can be avoided in the discrete model by considering appropriate velocity distributions $f(v_i;\tau_i)$ with finite support $[-c,c]$. To keep subsequent formulae as simple as possible, we shall use natural units with $c=1$ (even though, $c$ will be reinstated occasionally in figure captions or to clarify units). Moreover, for simplicity, we focus on the case of an equally spaced partition
\be\label{e:partition}
\tau_i\equiv \tau=t/N,\qquad \forall\; i=1,\ldots,N,
\ee
and write $f(v_i)$ instead of $f(v_i;\tau)$.

\subsubsection{Time-discrete Markov model}
\label{relativistic-a}
The relativistic generalization of the one-dimensional, nonrelativistic Maxwell-distribution~\eqref{e:Maxwell} is given by the one-dimensional modified J\"uttner function~\cite{DuHa07,DuTaHa07b}
\be\label{e:Juettner}
f(v_i)=
\f{\gc(v_i)^{2}}
{2\,K_0(\chi)}\;\exp\bigl[-\chi\gc(v_i)\bigr]\;
\Gt(1-|v_i|),
\ee
where $\chi$ is a (dimensionless) temperature parameter, and $K_0(\chi)$ a modified Bessel-function of the second kind~\cite{AbSt72}. Furthermore,
$$
E=m\gc(v),\quad 
p=m\gc(v)\;v, \quad
\gc(v)=(1-v^2)^{-1/2}
$$
denote the relativistic kinetic energy, momentum and $\gc$-factor, respectively. Analogous to the Maxwell distribution in the nonrelativistic case, the PDF~\eqref{e:Juettner} is distinguished by the fact that it is conserved in relativistic elastic binary collisions of classical particles~\cite{DuHa07}. 
\par
From the nonrelativistic limit, i.e., by expanding the exponent of Eq.~\eqref{e:Juettner} for $v_i^2\ll 1$ and comparing with Eq.~\eqref{e:Maxwell}, one can identify
\be\notag
\chi=m\gb=\f{m}{\kB T}=\f{\tau}{2D}.
\ee
Upon inserting the PDF~\eqref{e:Juettner} into Eq.~\eqref{e:propagator}, one obtains 
\be
p_N(t,x|x_0)
&=&\notag
\left\{
\prod_{i=1}^N\int \diff v_i\,
\f{\gc(v_i)^{2}}{2K_{0}\bigl[\tau/(2 D)\bigr]} 
 \right\}
\times\\
&& \notag\qquad
\exp\biggl(-\sum_{i=1}^N\f{\tau\, \gc(v_i)}{2D}\biggr)
\times\\
&&\qquad\label{e:relativistic_propagator_N}
\gd\biggl((x-x_0)-\sum_{i=1}^{N} \tau\,v_i\biggr).
\ee
In contrast to the nonrelativistic case, it is very difficult or, perhaps, even impossible to evaluate the integral~\eqref{e:relativistic_propagator_N} analytically for $N>1$. The propagator for the trivial case $N=1$  is given by
\be\label{e:rel-propagator_1}
p_1(t,x|x_0)
&=&\notag
\int_{-1}^{1}\diff v_1\; f(v_1)\,
\gd\bigl(x-x_0+\tau v_1\bigr)\\
&=&\notag
\f{1}{2 \tau K_0(\chi)}
\biggl[1-\f{(x-x_0)^2}{\tau^2}\biggr]^{-1}\times\\
&&\quad\notag 
\exp\biggl\{-\chi
\biggl[1-\f{(x-x_0)^2}{\tau^2}\biggr]^{-1/2}\biggr\}\;\times\\
&&\qquad 
\Gt(\tau-|x-x_0|).
\ee
Before discussing multiple scatterings $N>1$ we note that, in principle, there are two ways of reading the Eq.~\eqref{e:partition}: If one fixes $t$ and increases $N$, then this corresponds to taking the continuum limit $\tau\to 0$. Alternatively, one can consider the time-of-free-flight $\tau$ as being fixed; then the discretization \eqref{e:partition} just means that $t$ is given by multiples of $\tau$. In the remainder of this subsection, we shall adopt the latter point of view, because this leads to a nontrivial time-discrete relativistic Markov model. By contrast, as discussed afterwards, the continuum limit $\tau\to 0$ fails to produce a useful diffusion model in the relativistic case.
\par
\begin{figure}[t]
\center
\epsfig{file=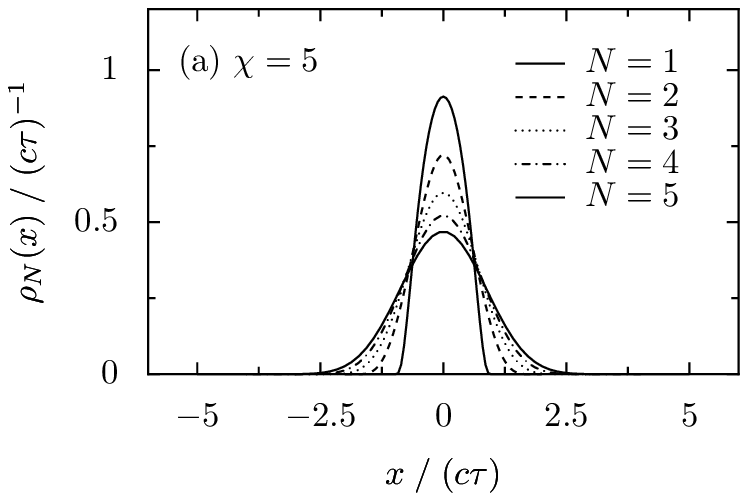}
\epsfig{file=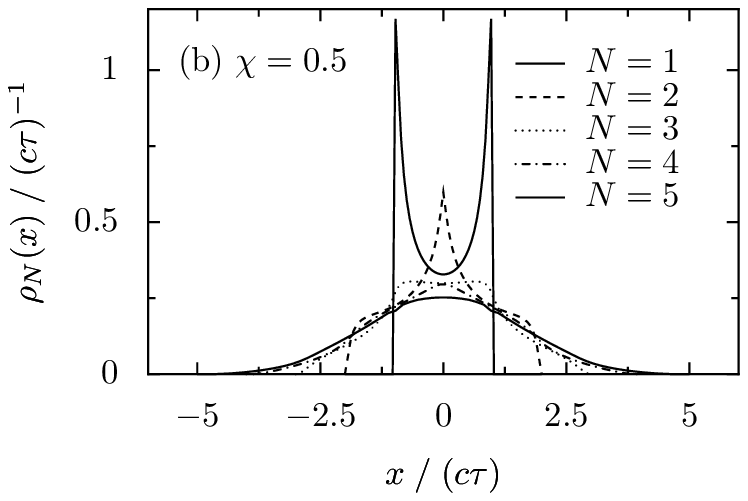}
\caption{Transition PDFs $\rho_N(x)=p_N(N\tau,x|0,0)$ for the relativistic time-discrete RW model from Sec.~\ref{relativistic-a}. The PDFs were numerically calculated at five different times $t=N\tau$ for two different parameter values $\chi=mc^2/(\kB T)=\tau/(2D)$. 
(a)~Weakly relativistic velocity distribution with $\chi> 1$. The numerical solutions look qualitatively similar to a (nonrelativistic) Gaussian, but have a finite support~$[-ct,ct]$. (b)~Strongly relativistic velocity distribution with $\chi< 1$. The peaks reflect the discrete time steps of the model, and are associated with the backward/forward-scattering of the fastest particles; cf. explanation in the text.
\label{fig02} }
\end{figure}
However, assuming for the moment that $\tau$ is fixed, the probability density at time $t=N\tau$ for a particle with initial condition $x_0=0$ is given by
$$
\rho_N(x):= p_N(N\tau,x|0).
$$  
Numerical results for different values of $N$ and two different parameter values $\chi=mc^2/(\kB T)$ or $D=\tau/(2\chi)$, respectively,  are depicted in Fig.~\ref{fig02}. As evident from the diagrams, superluminal particle propagation does not occur anymore, due to the choice of the velocity PDF~\eqref{e:Juettner}. Moreover, as the comparison with Fig.~\ref{fig01} shows, at large times $t\gg \tau$ the results start to look similar to those of the corresponding nonrelativistic diffusion process.

\par
Furthermore, for a strongly relativistic velocity PDF with \mbox{$\chi\lesssim 1$}, see Figs.~\ref{fig02} (b), we also observe peculiar peaks due to the discrete time structure of our model. Somewhat miraculously, such effects are absent in the nonrelativistic model (cf. Sec.~\ref{nonrelativistic}), but this is attributable to the fact that the Maxwellian (Gaussian) PDF represents a stable distribution~\cite{Feller2}; i.e., it is form-invariant under consecutive foldings as those in Eq.~\eqref{e:propagator}. In general, however, this type of invariance does not hold anymore if the velocity distribution is non-Gaussian and of finite support. In the latter case, characteristic relaxation patterns (peaks, dints, etc.) may arise which reflect the space/time discretization scheme underlying the RW model.
\par
For instance, the appearance and positions of the peaks in Figs.~\ref{fig02}~(b) can be readily understood  by considering the ultra-relativistic case $\chi\to 0$, where the velocity distribution~\eqref{e:Juettner} converges to two $\gd$-peaks at $\pm c$: Let us assume that an ultra-relativistic particle starts at time $t=0$ at position $x_0=0$. Then at time $t=\tau$ we have $p(x,\tau|0,0)=[\gd(x+\tau)+\gd(x-\tau)]/2$, corresponding to two peaks of equal height. Analogously, at  $t=2\tau$ one finds $p(x,2\tau|0,0)=[\gd(x+2\tau)+2\gd(x)+\gd(x-2\tau)]/4$ corresponding to three peaks, etc..
\par
From the general point of view, however, it would be desirable to also have continuous, analytically tractable diffusion models at one's disposal which do not exhibit such peculiar relaxation effects and discretization signatures. Such a model will be proposed in Sec.~\ref{axiomatic}. Before, it is worthwhile to study in more detail why the continuum limit $\tau\to 0$  does not provide a useful Markov model in the relativistic case.

\subsubsection{Shortfall of the relativistic continuum limit}
\label{shortfall}
Keeping $t$ fixed and taking the limit $N\to \infty$ of the time-discrete propagator~\eqref{e:relativistic_propagator_N}, one can write the relativistic analogon of the path integral~\eqref{e:nonrelativistic_propagator_pi} as 
\bse
\be\label{e:relativistic_propagator_pi}
p(t,x|x_0)&=&\notag
\int\D[V]\;\exp\biggl[-\int_0^t\diff s\; \f{\gc[V(s)]}{2D}\biggr]
\times\\
&&\qquad
\gd\biggl((x-x_0)-\int_0^t\diff s\; V(s)\biggr),
\qquad
\ee 
where
\be
\int\D[V]:=\lim_{N\to\infty}
\left\{
\prod_{i=1}^N\int \diff v_i\,
\f{\gc(v_i)^{2}}{2K_0\bigl[t/(2 DN)\bigr]} 
 \right\}.
\ee
\ese
However, as we shall see immediately, this implies the trivial result 
\be
p(t,x|x_0)=\gd(x-x_0),\qquad \fa t\ge 0.
\ee
The proof can be performed most easily by means of the Central Limit Theorem (CLT)~\cite{Ch01,JaPr03}. For this purpose, consider the mean velocity along a path, defined by 
\be\notag
\bar{V}_N(t):=\f{X(t)-x_0}{t}
\overset{\eqref{e:basic_equation_2a}}{=}
\f{1}{t} \sum_{i=1}^N \tau_i v_i
\overset{\eqref{e:partition}}{=}
\f{1}{N} \sum_{i=1}^N v_i.
\ee
The random variables $v_i$ are identically, independently distributed, with mean $\lan v_i\ran=0$ and variance \mbox{$\lan v_i^2\ran<\infty$}. For this case, the CLT asserts that the distribution of the random variable $Z_N(t):=\sqrt{N}\bar V_N(t)$ converges to a Gaussian. Hence, in the continuum limit $N\to \infty$, the mean velocity $\bar V_N(t)$ goes to zero with probability one, which means that the particle effectively remains at its initial position, i.e., $p(t,x|x_0)=\gd(x-x_0)$ as anticipated above.
\par
One may therefore conclude that a nontrivial continuous, relativistic Markov process in position space cannot exist~\footnote{A more elaborate proof is given by Dudley~\cite{Du65}.}. The main reason is that for \emph{any} relativistic velocity  distribution the variance is bounded $\lan v_i^2\ran<\infty$ because the values $v_i$ are bounded, and thus the CLT applies. This also means that, in principle, it is not very promising to consider  relativistic path integrals of the Markovian type~\eqref{e:relativistic_propagator_pi}. In particular, in view of the close analogy between diffusion and quantum propagators in the nonrelativistic case, this seems to leave very little, if any, room for formulating a relativistic particle quantum mechanics in terms of imaginary time Markov path integrals~\cite{Kleinert}.

\section{Non-Markovian continuous diffusion models}
\label{axiomatic}

The preceding discussion has shown that one has to abandon the Markov 
property if one wishes to define a continuous diffusion model in 
position space that avoids superluminal particle propagation speeds. 
In this section, we identify a simple \emph{non}-Markovian generalization 
of the Wiener propagator~\eqref{e:solution_non-rel} that is confined to the 
light cone and exhibits continuous diffusion fronts (in contrast to the 
solution of the telegraph equation). 
As we shall see, the specific functional shape of the 
proposed relativistic propagator naturally arises from 
the nonrelativistic case, provided one rewrites the transition 
PDF of the Wiener process in terms of an integral over actions. 
\par
We generalize the discussion to an arbitrary number of space dimensions~$d$ from now on.
Abbreviating $\bar x=(t,\bs x)$ and $\bar x_0=(t_0,\bs x_0)$, the 
transition PDF of the $d$-dimensional Wiener process is given by
\be\label{e:solution_non-rel-d}
p(\bar x|\bar x_0)
&=&\mcal{N}_d\,
\exp\biggl[-\f{(\bs x-\bs x_0)^2}{4D(t-t_0)}\biggr],
\ee 
where the normalization constant $\mcal{N}_d$ is a function of
the time difference $(t_1-t_0)>0$. Analogous to the one-dimensional case
 from Sec.~\ref{nonrelativistic}, the Wiener 
propagator~\eqref{e:solution_non-rel-d} could be expressed via 
a standard Markovian path-integral construction. But again, this type
of representation is of no use with regard to the relativistic 
generalization for the same reasons as discussed in 
Sec.~\ref{shortfall}. Hence, we shall next look for another 
formal representation of Eq.~\eqref{e:solution_non-rel-d}, that can 
be consistently transferred to the relativistic case.
\par
For this purpose, consider a particle traveling from the event
$\bar x_0=(t_0,\bs x_0)$ to $\bar x=(t,\bs x)$. Assume that the  
particle can experience multiple scatterings on its way, and
that the velocity is approximately constant between two successive 
scattering events. Then the total action (per mass) required along the path 
is given by
\be\label{e:action_nonrelativistic}
a=\f{1}{2}\int_{t_0}^t\diff t'\;\bs v(t')^2,
\ee
where the velocity $\bs v(t')$ is a piecewise constant function.
Clearly, the action becomes minimal for the deterministic (direct) 
path, i.e., if the particle does \emph{not} collide at all. In this case
it moves with constant velocity 
\mbox{$\bs v(t')\equiv(\bs x-\bs x_0)/(t-t_0)$} for 
all $t'\in[t_0,t]$, yielding the minimum action value
\be\label{e:action_1}
a_-(\bar x,\bar x_0)=\f{1}{2}\f{(\bs x-\bs x_0)^2}{t-t_0}.
\ee
On the other hand, to match the boundary conditions it is merely required that the mean velocity equals $(\bs x-\bs x_0)/(t-t_0)$.  Consequently, in the nonrelativistic case, the absolute velocity of a particle may become arbitrarily large in some intermediate time interval $[t',t'']\subset [t_0,t]$. Hence, the largest possible action value is \mbox{$a_+(\bar x,\bar x_0)=+\infty$}.
These considerations put us in the position to rewrite the Wiener 
propagator~\eqref{e:solution_non-rel-d} as an integral over actions:
\be\label{e:action-rep}
p(\bar x|\bar x_0)&=&
\mcal{N}_d'
\int_{a_-(\bar x|\bar x_0)}^{a_+(\bar x|\bar x_0)} 
\diff a\; \exp\biggl(-\f{a}{2D}\biggr).
\qquad
\ee 
The important advantage of this representation lies in the fact that
it may be generalized to the relativistic case in a straightforward
manner: One merely needs to insert the corresponding relativistic 
expressions into the boundaries of the integral.
\par
The relativistic generalization of 
Eq.~\eqref{e:action_nonrelativistic} reads~\cite{Weinberg}
\be\label{e:action_relativistic}
a=-\int_{t_0}^t\diff t'\sqrt{1-\bs v(t')^2}.
\ee
Analogous to the nonrelativistic case, the relativistic 
action~\eqref{e:action_relativistic} assumes its 
minimum $a_-$ for the deterministic (direct) path 
from $\bs x_0$ to $\bs x$, characterized by a constant velocity 
$\bs v(t')\equiv(\bs x-\bs x_0)/(t-t_0)$. One explicitly obtains
\bse\label{e:rel_PDF_best}
\be\label{e:action_2}
a_-(\bar x,\bar x_0)
=-\left[(t-t_0)^2-(\bs x-\bs x_0)^2\right]^{1/2},
\ee 
i.e., $a_-$ is the negative Minkowski distance 
of the two space-time events $\bar x_0$ and $\bar x$.
The maximum action value is realized for particles 
moving at light speed, yielding $a_+=0$.  
Hence, the transition PDF for the relativistic generalization 
of the Wiener process reads
\be\label{e:rel_PDF_new}
p(\bar x|\bar x_0)=
\mcal{N}_d\;\left\{\exp\biggl[-\f{a_-(\bar x,\bar x_0)}{2D}\biggr]
-1\right\},
\ee
\ese
if $(\bs x-\bs x_0)^2\le (t-t_0)^2$, and $p(\bar x|\bar x_0)\equiv 0$ 
otherwise, with $a_-$ determined by Eq.~\eqref{e:action_2}.
Explicit results for the normalization constant $\mcal{N}_d$ with $d=1,2,3$ 
are given in App.~\ref{a:normalization}. 
\par
\begin{figure}[t]
\center
\epsfig{file=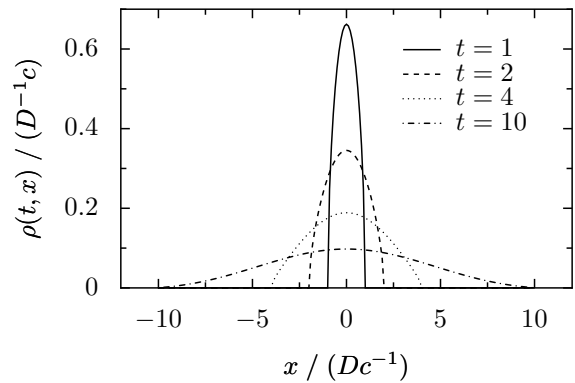}
\caption{Transition PDF $\rho(t,x)= p(t,x|0)$ for the one-dimensional ($d=1$) relativistic diffusion process~\eqref{e:rel_PDF_best} at different times $t$ (measured in units of $D/c^2$).
\label{fig03}}
\end{figure}
\begin{figure}[t]
\center
\epsfig{file=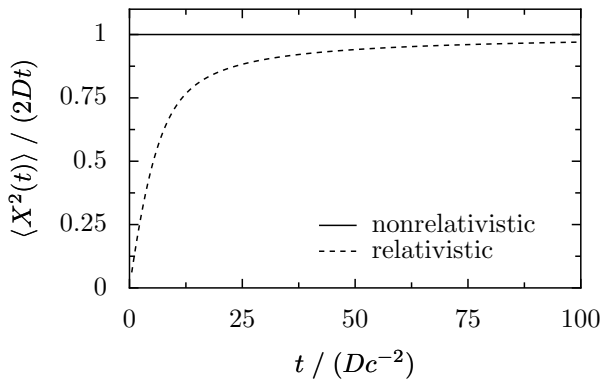}
\caption{Comparison of the mean square displacements $\lan X^2(t)\ran$, divided by $2Dt$, for the one-dimensional (\mbox{$d=1$}) nonrelativistic Wiener process~\eqref{e:solution_non-rel} and its relativistic generalization from Eq.~\eqref{e:rel_PDF_best} with initial condition $(t_0,x_0)=(0,0)$. 
\label{fig04}
}
\end{figure}

Figure~\ref{fig03} shows the PDF $\rho(t,x)= p(t,x|0,0)$ of the 
diffusion process~\eqref{e:rel_PDF_best} for the one-dimensional 
case \mbox{$d=1$} at different times $t$. The corresponding mean 
square displacement is plotted in Fig.~\ref{fig04} (dashed curve). 
\par
The diffusion process defined by Eqs.~\eqref{e:rel_PDF_best} is \emph{non-Markovian}. 
This can be seen, e.g., by testing the necessary Chapman-Kolmogorov criterion  
\be\label{e:Markov-condition}
p(t,\bs x|t_0, \bs x_0)=
\int_{\R^d} \diff^d \bs x_1\;p(t,\bs x|t_1,\bs x_1)\;
p(t_1,\bs x_1|\bs x_0).
\quad
\ee
If Eq.~\eqref{e:Markov-condition} becomes invalid for some $t_1\in (t_0,t)$, 
then the process is non-Markovian. As one my easily check, e.g., by 
numerically evaluating~\eqref{e:Markov-condition} for some sample 
values $(t,\bs x;t_0,\bs x_0)$ and some intermediate time $t_1$, 
Eq.~\eqref{e:Markov-condition} is indeed violated for the transition 
PDF~\eqref{e:rel_PDF_best}.  
\par
It is interesting to note that the PDF~\eqref{e:action-rep} 
is a special case of a larger class of diffusion processes, defined by
\be\label{e:embedding}
p_w(\bar x|\bar x_0)&=&
\mcal{N}_{w}
\int_{a_-(\bar x|\bar x_0)}^{a_+(\bar x|\bar x_0)} \diff a\; w(a),
\qquad
\ee 
where $w(a)\ge 0$ is a weighting function, and $\mcal{N}_{w}$ the 
time-dependent normalization constant. 
Equation~\eqref{e:action-rep} is recovered by 
choosing~$w(a)\equiv \exp[-a/(2D)]$. 
It is, however, worth mentioning that a very large class of functions $w(a)$ 
yields an asymptotic growth of the spatial mean square displacement 
that is proportional to $t$, corresponding to \lq ordinary\rq\space 
diffusion. Moreover, Eq.~\eqref{e:embedding} can also be used to 
describe superdiffusion or subdiffusion processes~\cite{1990BoGe,1999Ca,2000MeKl}, whose asymptotic 
mean square displacements grow as $t^\ga,\;\ga\ne 1$~\footnote{This 
can be achieved, e.g., by choosing the integral 
boundaries as $\tilde a_-= (\bs x-\bs x_0)^2/(t-t_0)^\ga,\;\ga\ne 1$ and 
$a_+=\infty$, but then the variable $a$ may not be 
interpreted as a conventional action anymore.}.
\par
In particular, Eq.~\eqref{e:embedding} may be viewed as 
an \lq unconventional\lq\space non-Markovian path integral definition in 
the following sense: Physically permissible paths from $\bar x_0$ to $\bar x$ 
have action values (per mass) $a$ in the range $[a_-,a_+]$. Grouping the 
different paths together according to their action values, one may assign 
to each such class of paths, denoted by $\mcal{C}(a;\bar x,\bar x_0)$, the 
statistical weight $w(a)$. The integral~\eqref{e:embedding} can then be 
read as an integral over the equivalence classes $\mcal{C}(a;\bar x,\bar x_0)$ 
and their respective weights~$w(a)$. The nonrelativistic Wiener process 
corresponds to the particular choice $w(a)=\exp[-a/(2D)]$; hence, it is 
natural to define the relativistic generalization by using the same weighting 
function. 

\section{Summary}

The search for a consistent description of diffusion processes in the framework of special relativity represents a longstanding issue in statistical physics. Several different relativistic generalizations of the nonrelativistic diffusion equation~\eqref{e:diffusion} have been proposed in the past~\cite{Ha65,MaWe96,BoPoMa99,HePa01,KoLi00,KoLi01,AbGa04,AbGa05}. The most prominent example is the telegraph equation~\eqref{e:telegraph}. This second order partial differential equation (PDE) describes a non-Markovian diffusion process without superluminal particle propagation~\cite{Ta22,Go50,RevModPhys.61.41,RevModPhys.62.375}. However, the solutions of the telegraph equation exhibit singular diffusion fronts propagating at light speed $c$, and may therefore be less appropriate for the description of massive particles. 
\par   
To avoid such singularities, we pursue a complementary approach in this paper; i.e., instead of trying to find a relativistic diffusion PDE we focus on the transition probability density function (PDF) of a relativistically acceptable diffusion model. As the main result, an analytic expression for the transition PDF is proposed in Eq.~\eqref{e:rel_PDF_best}. The simple relativistic diffusion process defined by this PDF avoids superluminal propagation, and reduces to the Wiener process~\eqref{e:solution_non-rel} in the nonrelativistic domain. This analytically tractable relativistic diffusion model can be useful for practical calculations in the future (e.g., in high energy physics \cite{AbGa04,AbGa05} and astrophysics~\cite{RyLi85}). 
\par
Finally, it is worth emphasizing that, analogous to the solutions of TE but in contrast to the Wiener process, the proposed diffusion model is non-Markovian. This is not a drawback, as it is generally impossible to construct nontrivial continuous relativistic Markov processes in position space (cf. discussion in Sec.~\ref{relativistic} and Ref.~\cite{Du65}). In view of the close analogy between diffusion and quantum propagators in the nonrelativistic case, it would be quite interesting to learn, if this also means that it is impossible to formulate a relativistic particle quantum mechanics on the basis of standard path integrals~\cite{Kleinert,St04}.

\begin{acknowledgments}
We would like to thank Stefan Hilbert for helpful discussions, and Mari\'an Bogu\~n\'a for bringing interesting references to our attention.
\end{acknowledgments}


\bibliography{TD,RelTD,Papers,RelDiff,RBM,ActiveBM,RBMapplied,FokkerPlanck,BrownianMotion,RelKin,Journals,Books,PhotonsDiffusion,QFT,MathBooks,PathIntegrals,Anomalous,Debbasch}

\appendix


\section{Calculation of the normalization constant $\mcal{N}_d$}
\label{a:normalization}

We wish to express the normalization constant $\mcal{N}_d$ from Eq.~\eqref{e:rel_PDF_new} in terms of modified Bessel functions of the first kind and modified Struve functions \cite{AbSt72}. Introducing $\mbox{$\bs z:=\bs x-\bs x_0$}$ and $u:=t-t_0$, we have to calculate
\be
\mcal{N}_d&=&\notag
\int_{\mbb{R}^d}\!\!\!
\diff^d\bs z\;
\Gt(u-|\bs z|)\,
\biggl\{
\exp\biggl[\f{(u^2-\bs z^2)^{1/2}}{2D}\biggr]
-1\biggr\}. 
\ee
Using spherical coordinates, we can rewrite this as 
\be\label{a-e:norm_2}
\mcal{N}_d&=&\notag
O_d\int_0^{u}\diff|\bs z |\; |\bs z |^{d-1}
\;\biggl\{
\exp\biggl[\f{(u^2-|\bs z|^2)^{1/2}}{2D}\biggr]
-1\biggr\},\\
\ee
where $O_d={2\pi^{d/2}}/{\Gc(d/2)}$ is the surface area of the $d$-dimensional unit-sphere. It is convenient to split the integral~\eqref{a-e:norm_2} in the form
\be
\mcal{N}_d&=&\mcal{N}_d'-\f{u^d}{d} O_d,
\ee
where
\be\label{a-e:norm_2a}
\mcal{N}_d'&=&
O_d\int_0^{u}\diff|\bs z |\; |\bs z |^{d-1}
\;
\exp\biggl[\f{(u^2-|\bs z|^2)^{1/2}}{2D}\biggr],\quad
\ee
Next we substitute \mbox{$|\bs z|= u\,\sin \phi$}, where $\phi\in[0,\pi/2]$. Then Eq.~\eqref{a-e:norm_2a} takes the form
\be
\mcal{N}_d'&=&\notag
u^d\;O_d\int_0^{\pi/2}\diff \phi\;\cos\phi\;\sin^{d-1}\phi\;
\exp\biggl(\f{u\,\cos \phi}{2D}\biggr).
\ee
Based on this integral representation, $\mcal{N}_d'$ can be expressed in terms of modified Bessel functions of the first kind $I_n$, and modified Struve functions $L_k$ \cite{AbSt72}, and one finds
\bse
\be
\mcal{N}_1'&=& u\;\pi\,
\left[I_1(\chi)+L_{-1}(\chi)\right],\\
\mcal{N}_2'&=& u^2\;\f{2\pi}{\chi^2}\,
\left[1+(\chi-1)\;\exp(\chi)\right],\\
\mcal{N}_3'&=&
u^3\;\f{2\pi^2}{\chi^2}\,
\left\{\chi\left[I_2(\chi)+L_0(\chi)\right]-2L_1(\chi)\right\},
\qquad
\ee
\ese
where $ \chi={u}/{(2D)}$.

\end{document}